\documentclass[letterpaper,english,aps,prb,floatfix,twocolumn,,showpacs,amsfonts,amssymb,superscriptaddress]{revtex4}
\usepackage[T1]{fontenc}
\usepackage[latin1]{inputenc}
\usepackage{graphics}
\usepackage{amssymb}
\usepackage{amsmath}
\usepackage{overpic}

\newcommand{\be}{\begin{equation}}
\newcommand{\ee}{\end{equation}}
\newcommand{\bea}{\begin{eqnarray}}
\newcommand{\eea}{\end{eqnarray}}

\def\a{\alpha}

\def\d{\delta}

\def\m{\mu}
\def\l{\lambda}

\def\s{\sigma}

\def\D{\Delta}
\def\O{\Omega}

\def\ra{\rightarrow}

\def\nn{\nonumber}
\def\lb{\label}
\def\pref#1{(\ref{#1})}

\newcount\bozza \bozza=0
\ifnum\bozza=1
\newdimen\shift \shift=-2truecm
\def\lb#1{%
{\label{#1}\rlap{\kern\shift{$\scriptstyle#1$}}}}
\else\def\lb#1{\label{#1}} \fi

\begin{document}

\title{Broadening of the Beresinkii-Kosterlitz-Thouless superconducting
  transition by inhomogeneity and finite-size effects}

\author{L.~Benfatto}
\affiliation{Centro Studi e Ricerche ``Enrico Fermi'',
via Panisperna 89/A, 00184,
Rome, Italy}
\affiliation
{CNR-SMC-INFM and Department of Physics, University of Rome ``La
  Sapienza'',\\ Piazzale Aldo Moro 5, 00185, Rome, Italy}

\author{C.~Castellani}

\affiliation
{CNR-SMC-INFM and Department of Physics, University of Rome ``La
  Sapienza'',\\ Piazzale Aldo Moro 5, 00185, Rome, Italy}

\author{T.~Giamarchi}
\affiliation{Universit\'e de Gen\`eve, DPMC, Quai Ernest-Ansermet CH-1211
Gen\`eve 4, Switzerland}

\date{\today}

\begin{abstract}
We discuss the crucial role played by finite-size effects and inhomogeneity
on the Beresinkii-Kosterlitz-Thouless (BKT) transition in two-dimensional
superconductors. In particular, we focus on the temperature dependence of
the resistivity, that is dominated by superconducting fluctuations above
the BKT transition temperature $T_{BKT}$ and by inhomogeneity below it. By
means of a renormalization-group approach we establish a direct
correspondence between the parameter values used to describe the BKT
fluctuation regime and the distance between $T_{BKT}$ and the mean-field
Ginzburg-Landau transition temperature. Below $T_{BKT}$ a resistive tail
arises due to finite-size effects and inhomogeneity, that reflects also on
the temperature dependence of the superfluid density. We apply our results
to recent experimental data in superconducting LaAlO$_3$/SrTiO$_3$
heterostructures, and we extract several informations on the microscopic
properties of the system from our BKT fitting parameters. Finally, we
compare our approach to recent data analysis presented in the literature,
where the physical meaning of the parameter values in the BKT formulas has
been often overlooked.
\end{abstract}


\maketitle

\section{Introduction}

In the last years a renewed interest emerged in the superconducting
transition in two-dimensional (2D) systems, prompted by the experimental
achievement of high-quality ultra-thin films of superconducting
materials. To this category belong both few-unit-cell thick films of
layered cuprate
superconductors\cite{martinoli_films,triscone_fieldeffect,lemberger_bilayer}
and the nanometer-thick layers of superconducting electron systems formed
at the interface between insulating oxides in artificial
LaAlO$_3$/SrTiO$_3$
heterostructures\cite{triscone_science07,triscone_nature08}. At the same
time, the experimental progresses made in the last decade of intense
investigation in high-temperature superconductors prompted additional
measurements in thin films of conventional superconductors by means of
different techniques or higher resolution than the ones available in the
past.  Typical examples are provided by the finite-frequency study of the
optical magneto-conductivity in films of InO$_x$\cite{armitage_films},
Nernst-effect measurements in amorphous films of NbSi\cite{lesueur_nerst}
and scanning tunneling microscopy in TiN
films\cite{sacepe_stm_prl08,sacepe_stm_cm09}.

Due to the 2D nature of these systems, the superconducting (SC) transition
is expected to belong to the Beresinskii-Kosterlitz-Thouless (BKT)
universality
class\cite{berezinsky,kosterlitz_thouless,kosterlitz_renormalization_xy},
where the gauge symmetry is unbroken in the SC state but the system has a
finite superfluid density, which is destroyed at $T_{BKT}$ by proliferation
of vortex-antivortex phase fluctuations. As it has been discussed at length
in the past literature, the BKT transition has in principle very specific
signatures\cite{review_minnaghen,giamarchi_book_1d}. For
example, by approaching the transition from below, the superfluid density
$n_s$ is expected to go to zero discontinuously at the BKT temperature
$T_{BKT}$, with an ``universal'' relation between $n_s(T_{BKT})$ and
$T_{BKT}$
itself\cite{nelson_univ_jump_prl77,review_minnaghen,giamarchi_book_1d}.
Approaching instead the transition from above one has in principle the
possibility to identify the BKT transition from the temperature dependence
of the superconducting fluctuations.  Indeed, in 2D the temperature
dependence of several physical quantities (like the paraconductivity or the
diamagnetism) is encoded in the temperature dependence of the
superconducting correlation length $\xi(T)$, that diverges exponentially at
$T_{BKT}$, in contrast to the power-law expected within Ginzburg-Landau
(GL) theory\cite{varlamov_book}.  A possible interpolation scheme between
standard GL fluctuations and BKT phase fluctuations of the SC
order parameter was proposed long ago in a seminal paper by Halperin and
Nelson\cite{halperin_ktfilms} (HN).

In the attempts made in the past to find out experimental signatures of the
BKT transition in thin films of conventional
superconductors\cite{fiory_prb83,kapitulnik_prb92,goldman_prb83} it turned
out that BKT fluctuations are usually restricted to a small temperature
regime near the GL transition temperature $T_{c}$. If the energy range
$T_c-T_{BKT}$ is extremely small, most of the fluctuation regime is
dominated by GL fluctuations, and the exponential signatures of BKT
fluctuations can be hardly detected. The predominance of the GL
fluctuation regime
has been confirmed also by more recent  
measurements of Nernst effect in NbSi films\cite{lesueur_nerst} and
zero-bias tunneling conductance in TiN
films\cite{sacepe_stm_cm09}. Analogously, the expected universal jump of
$n_s(T_{BKT})$ at $T_{BKT}$ due to vortex proliferation can be overscreened
by the simultaneous fast decrease of $n_s(T)$ due to quasiparticle
excitations near $T_c$\cite{goldman_prb83}.

An additional effect that can mask the occurrence of BKT transition is the
intrinsic inhomogeneity of the sample. For example, as it has been
discussed recently in the context of thin films of high-temperature
superconductors\cite{benfatto_kt_bilayer}, the spatial inhomogeneity can
broaden considerably the universal jump of the superfluid density, leading
to a smooth downturn of the superfluid density instead of the sharp one
expected in ultra-thin samples. Such an intrinsic mesoscopic inhomogeneity
has been revealed by scanning tunneling spectroscopy in cuprate
superconductors\cite{yazdani_07}, and recently also in films of
conventional supercondutcors\cite{sacepe_stm_prl08}. This indicates that
inhomogeneity is a crucial ingredient to several superconducting systems in
the presence of disorder, as suggested also by recent numerical
simulations\cite{dubi_nature07}.  Finally, one must account also for
finite-size effects (at the scale of the sample dimensions or even smaller,
in the inhomogeneous case), that are expected to cut off the divergence of
the correlation length at $T_{BKT}$.

In this work we aim to address the role of inhomogeneity and finite-size
effects in the BKT transition, providing a general scheme for analyzing
paraconductivity measurements in quasi-2D superconductors.  We first match
the HN interpolation scheme with a detailed analysis of the renormalization
group (RG) equations for the BKT transition far from the critical region
where an analytical solution is
available\cite{kosterlitz_renormalization_xy}, taking into account
finite-size effects. This analysis allows us to relate the parameters of
the standard BKT correlation-length expression to microscopic quantities,
reducing considerably the degrees of freedom in the fitting procedure of
the resistivity data.  In particular, we show that once fixed the
difference $T_c-T_{BKT}$ and the value of the vortex-core energy, whose
role has been recently discussed in the context of the physics of
high-temperature
superconductors\cite{benfatto_kt_rhos,benfatto_kt_magnetic,altman_prl07,benfatto_kt_bilayer,huse_prb08},
the behavior of the resistivity from $T_{KT}$ up to temperatures far above
$T_c$ is uniquely determined. In this way we establish a 
consistency check that the fitting parameters must satisfy when 
the standard BKT approximated
formulas are used, a fact that has been often overlooked in the
literature. 
Building on such an analysis we can also
take into account the role of inhomogeneities, which as we will
show are crucial to understand the experimental situation.
Indeed even relatively small inhomogeneities can provide a
significant enhancement of the resistive response below the
$T_{BKT}$ transition, that is simultaneously reflected in the
temperature dependence of the superfluid density across the
transition itself. A paradigmatic example of application of our
analysis is provided by recent measurements in superconducting
heterostructures\cite{triscone_science07,triscone_nature08}. As
we shall see, a correct treatment of finite-size effects and
inhomogeneity allows us to reproduce with great accuracy the
available experimental results, and to estimate the superfluid
density in these unconventional systems.

\begin{figure}[htb]
\includegraphics[scale=0.5,angle=0]{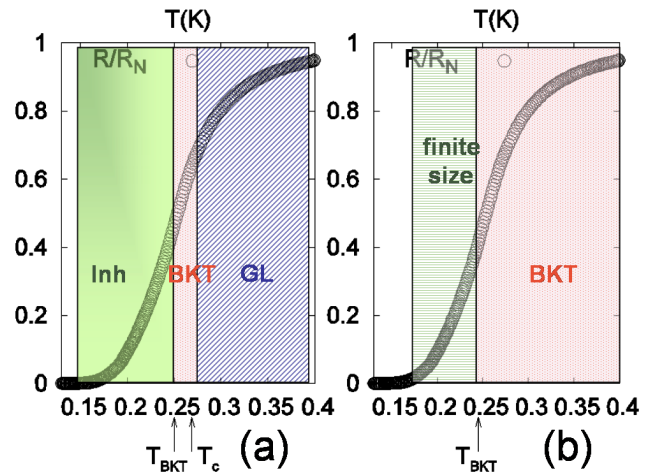}
\caption{(Color online) Comparison between the approach discussed in our
  paper and the one presented so far in the
  literature\cite{triscone_science07,triscone_nature08,schneider_finitesize09}
  as far as the resistivity data in superconducting heterostructures are
  concerned. The experimental data for the resistivity $R$ normalized to
  the normal-state value $R_N$ are taken from Ref.\
  \onlinecite{triscone_nature08}, and are the same as the ones showed in
  Fig.\ \ref{fig_nature} and Fig.\ \ref{fig_schneider} below. On the left
  panel we summarize our approach: most of the fluctuation regime above
  $T_{BKT}$ is dominated by GL fluctuations, while BKT fluctuations are
  restricted to a narrow range of temperatures near the transition
  temperature $T_{BKT}$ that would be observable in the homogeneous
  system. However, inhomogeneity leads to a considerable tail in the
  resistivity, that remains finite below $T_{BKT}$. On the right panel we
  summarize the point of view used in Ref.\
  \onlinecite{triscone_science07,triscone_nature08,schneider_finitesize09}:
  the whole range of temperatures above $T_{BKT}$ is dominated by SC
  fluctuations having BKT character, and finite-size effects are
  responsible for the resistive tail below $T_{BKT}$.}
\label{fig_scheme}
\end{figure}

We note that although the BKT transition has been already invoked to
discuss the physics of such heterostructures, a very different approach was
taken so far in the
literature\cite{triscone_science07,triscone_nature08,schneider_finitesize09},
relying essentially on the idea that the GL temperature $T_c$ is far larger
than $T_{BKT}$ (that can have eventually a different qualitative
character\cite{triscone_science07}), so that the \emph{whole fluctuation
  regime} can be described within BKT theory. A schematic view of the
difference between our analysis and the previous approach is shown in
Fig.\ \ref{fig_scheme}.  The point of view taken in
Refs.\ [\onlinecite{triscone_science07,triscone_nature08,schneider_finitesize09}]
leads of course to a very different identification of the various
temperature regimes, and very different physical parameters for the
underlying BKT theory. In our view these previous analyses suffer from
several problems, that we will discuss explicitly in this paper. In
particular they do not correctly identify the physical parameters of the
system and thus overlook several interesting consequences that one can
extract from the accurate experimental measurements performed in these new
materials.

The structure of the paper is the following. In Sec.~\ref{sec:fluct} we
review the standard description of fluctuation conductivity in 2D
superconductors, that establishes the correspondence between resistivity
and the fluctuation correlation length. In Sec.~\ref{sec:correl} we analyze
systematically the behavior of the correlation length within BKT theory, by
means of a RG approach. This allows us to fully identify the parameters in
the HN formula interpolating between the BKT transition and the standard
Gaussian fluctuations. Sec.~\ref{sec:inhom} is devoted to the discussion of
the role of inhomogeneity, trough the direct application of our
interpolating BKT-to-GL scheme to the resistive transition.  We apply our
scheme to identify the relevant parameters in superconducting
heterostructures. In Sec.~\ref{sec:discuss} we clarify the differences
between our approach and previous theoretical reports. The reader
interested only to the issue of analyzing the experimental
data can refer directly to the Sections \ref{sec:inhom} and
\ref{sec:discuss}. Finally, Sec.~\ref{sec:concl} contains the concluding
remarks.

\section{Fluctuation conductivity in 2D superconductors} \label{sec:fluct}

The definition of fluctuation conductivity within the BKT theory relies on
the Bardeen-Stephen formula\cite{tinkham_book}, which gives the excess
conductivity $\Delta \sigma\equiv \sigma-\sigma_N$ with respect to the
normal-state conductivity $\sigma_N$ as a function of the density of 
free vortices $n_F$ above $T_{BKT}$ as: 
\be
\Delta \s=\frac{e^2}{\hbar^2\pi^2 \mu_V}\frac{1}{n_F}
\ee
where the vortex mobility is $\mu_V=2\pi \xi_0^2 c^2 \rho_n/\Phi_0^2$, $\xi_0$
is the spacing for the vortex lattice, that we will assume equal to the
zero-temperature coherence length, $\rho_n$
is the normal-state resistivity and $\Phi_0$ the flux quantum. The vortex
density is also conventionally defined in terms of the correlation length
$\xi$ as $2\pi n_F\equiv 1/\xi^2$, so that near the transition where $\D
\s\gg \s_N$ the ratio between the resistance $R$ and its normal-state value
$R_N$ is:
\be
\lb{rnf}
\frac{R}{R_N}=2\pi \xi_0^2 n_F=\left(\frac{\xi_0}{\xi}
\right)^2,
\ee
which is the formula usually quoted in the literature for the
paraconductivity due to vortices, once that $\xi(T)$ is calculated within
BKT theory. Remarkably, in 2D the same formula is valid for the
Aslamazov-Larkin contribution of GL
fluctuations\cite{aslamazov_larkin,varlamov_book}. However, in this case
the temperature dependence of $\xi(T)$ follows $\xi_{GL}^2\sim 1/\log
(T/T_{c})$, or equivalently $\xi_{GL}^2\sim T_{c}/(T-T_{c})$, where $T_c$
is the GL or mean-field critical temperature. In 2D films the true
transition occurs at a $T_{BKT}$ lower than $T_c$, the distance between the
two being a function of the microscopic parameters of the system. Thus, the
fluctuation conductivity crosses over from a GL regime, where it shows a
tendency to a power-law divergence at $T_c$, to a KT regime, where the
correlation length diverges asymptotically when $T\ra T_{BKT}$ as $\xi\sim
\exp(b/\sqrt{t})$, where we introduced the reduced temperatures:
\be
\lb{red}
t\equiv \frac{T-T_{BKT}}{T_{BKT}}, \quad
t_c\equiv \frac{T_c-T_{BKT}}{T_{BKT}}.
\ee
As it was observed already by Halperin and
Nelson\cite{halperin_ktfilms}, the BKT fluctuations are
expected to be present only in the range of temperatures  $t\ll
t_c$. Thus, using the general expression $\D \s\propto \xi^2$,
they proposed a well-known interpolation formula between the GL
and BKT regimes given by:
\be
\lb{shn}
\D\sigma=a_{hn}\s_N \sinh^2(\sqrt{b_{hn} t_c/t})
\ee
where $b_{hn}$ is a dimensionless constant of order one. Thus, for
$b_{hn}t_c/t\gg 1$ one recognizes the exponential divergence characteristic
of the BKT theory, while for $b_{hn}t_c/t \ll 1$ one recovers a power-law
increase typical of GL fluctuations. For the prefactor $a_{hn}$ Halperin
and Nelson assumed $a_{hn}=0.37/b_{hn}$, following the
Beasley-Mooij-Orlando\cite{beasley_prl79} approximate relation between
$\s_N$ and the ratio $T_{BKT}/T_c$ in dirty BCS superconductors. We note in
passing that in the dirty BCS limit one can indeed consider only the AL
contribution to the GL paraconductivity, while in clean samples also the
Maki-Thomson contribution can be sizable, making the analysis of GL
fluctuations more involved\cite{kapitulnik_prb92}.

In the HN formula some ambiguity is still present in the choice
of the parameters, i.e. the prefactor and the exponential
coefficient. In this work we want to fix this ambiguity by
determining exactly their values from a RG analysis of the BKT
correlation length. In this way, we provide a clear procedure
to analyze experimental data by respecting the internal
consistency between the fitting parameters for the
paraconductivity. Moreover, we will discuss how finite-size
effects and inhomogeneity affect the fluctuations resistivity
above and below $T_{BKT}$. All these issues turn out to be
crucial to correctly interpret the experimental data, as we
will discuss in Sec.~\ref{sec:inhom} and ~\ref{sec:discuss}.

\section{Behavior of the correlation length within the BKT
theory} \label{sec:correl}

As it was shown by Kosterlitz in its original
work\cite{kosterlitz_renormalization_xy,review_minnaghen}, the
critical properties of the BKT transition can be captured by
the analysis of the RG equations for the two main quantities
involved in the SC transition: the superfluid stiffness $J$ and
the vortex fugacity $g=2\pi e^{-\beta\mu(T)}$ ($\beta=1/k_B
T$), where $\mu$ is the vortex-core energy. The stiffness
$J$ is the energy scale associated to the 2D
superfluid density $n_s^{2D}$ (measured experimentally via the
inverse penetration depth $\lambda$), including
already the temperature depletion due to quasiparticle
excitations:
\be
\lb{defk}
\quad J(T)=\frac{\hbar^2 n_s^{2D}(T)}{4 m^* k_B}=
\frac{\hbar^2 c^2 }{16\pi e^2}\frac{d}{\l^2(T)},
\ee
where $d$ is the film thickness and $m^*$ is the effective mass of the
carriers. The
vortex-core energy $\mu$ is in general a multiple of the stiffness
itself\cite{nagaosa_book,benfatto_kt_rhos,benfatto_kt_bilayer}:
\be
\lb{defmu}
\mu(T)=\tilde \mu J(T)
\ee
where $\tilde \mu$ is a dimensionless constant. As we discussed
recently\cite{benfatto_kt_rhos,benfatto_kt_magnetic,benfatto_kt_bilayer}, a
general approach to the BKT transition can require to assume that
$\tilde \mu$ deviates with respect to the conventional value $\tilde
\mu_{XY}=\pi^2/2$ that it acquires
in the XY
model\cite{kosterlitz_renormalization_xy,review_minnaghen,nagaosa_book}, so
that we shall use
\be
\lb{def_mu_scale}
\tilde \mu=\alpha \tilde \mu_{XY}=\alpha \frac{\pi^2}{2}
\ee
The RG equations in the variables $g$ and $K=\pi J_s/T$ can be written (in
analogy with the notation used for the sine-Gordon
model\cite{giamarchi_book_1d}) as:
\bea
\lb{eqk}
\frac{dK}{d\ell}&=&-K^2g^2,\\
\lb{eqg}
\frac{dg}{d\ell}&=&(2-K)g,
\eea
where $\ell=\ln a/\xi_0$, and $\xi_0,a$ are the original and rescaled RG
lattice spacing, respectively.
The physical value of the superfluid density $J_s$ is
determined by the limiting value of $K$ under RG flow, i.e.
$J_s\equiv T K(\ell\ra \infty)/\pi$. In the low-temperature
regime $K>2$ so that the vortex fugacity scales to zero under
RG flow and $J_s$ is finite, with a small renormalization with
respect to the initial value. Instead at high temperature $K<2$
the vortex fugacity becomes relevant, it diverges under RG flow
and as a consequence $J_s$ scales to zero.  The BKT temperature
is the one where the above system of equations reaches the
fixed point $K=2, g=0$, so that at $T_{BKT}$
\begin{equation}
\lb{jump}
\frac{\pi J_s(T_{BKT})}{T_{BKT}} = 2,
\end{equation}
i.e. one recovers the universal jump of the superfluid
density\cite{nelson_univ_jump_prl77,review_minnaghen,giamarchi_book_1d}.

The usual definition\cite{kosterlitz_renormalization_xy,review_minnaghen}
of the correlation length $\xi$ above $T_{BKT}$ relies on the determination
in the RG equations \pref{eqk}-\pref{eqg} of a characteristic scale $\bar
\ell$ at which the vortex fugacity is ``sufficiently large''. The exact
definition of this scale is somehow arbitrary, but it does not change
qualitatively the results. In practice, we shall use as a working
definition the scale $\ell_s$ where the RG parameter $K(\ell_s)$ related to
the superfluid density vanishes. The vortex density and correlation length
are then defined as:
\be
\lb{defnv}
2\pi n_F=\frac{g(\bar\ell)}{2\pi a^2(\bar \ell)}=\frac{1}{\xi^2}, \quad
a(\bar \ell)=\xi_0e^{\bar \ell},
\ee
where $\bar \ell=min(\ell_s,\ell_{max})$, where $\ell_{max}$
takes into account finite-size effects, as we shall discuss
below. Near $T_{BKT}$ one expects to recover the
well-known\cite{kosterlitz_renormalization_xy}
exponential behavior of $\xi(T)$, that we will parameterize as:
\be
\lb{xiapp}
\frac{\xi}{\xi_0}=\frac{1}{A} e^{b/\sqrt{t}}, \quad t\ra 0
\ee
where $b$ and $A$ are constants of order one.

It should be emphasized that if the bare superfluid stiffness were a
constant independent on the temperature Eq.~\pref{xiapp} would be valid
until $t\sim {\cal O}(1)$, i.e. essentially at all relevant temperatures
above $T_{BKT}$. However, what limits in a crucial way the applicability of
the above approximation is the temperature dependence of $J(T)$ due to
quasiparticle excitations, which lead to the vanishing of $J(T)$ at the GL
temperature $T_c$. In a s-wave BCS superconductor a good approximation for
$J(T)$ is:
$$
\frac{J(T)}{J_0}=\left(\frac{\D(T)}{\D_0}\right)^2, \quad \D(T)=\D_0\tanh
\left(\frac{\pi}{2}\sqrt{\frac{T_c}{T}-1}\right).
$$
For any practical purpose, to determine $t_c$ what matters is
only the behavior of $J(T)$ near $T_c$, that according to the
above equation is linear:
\be
\lb{japp}
J(T)\approx J_0\frac{\pi^2}{4}\left(1-\frac{T}{T_c}\right), \quad T\approx T_c.
\ee
By neglecting the normalization of $J_s$ with respect to $J$
due to vortices already below $T_{BKT}$, one can approximately
estimate the BKT temperature by the condition \pref{jump} $J(T_{BKT})=2
T_{BKT}/\pi$, so that one gets:
\be
\lb{tcest}
t_c\approx \frac{8 T_c}{\pi^3 J_0}.
\ee
Thus, one sees that the larger is $J_0/T_c$ the smaller is the
interval $t_c$. In a thin film of conventional
superconductors\cite{fiory_prb83,kapitulnik_prb92,goldman_prb83}
the typical mean-field temperatures are of order of few K,
while $J_0$ can be as large as the Fermi energy if
$n_s^{2D}(0)$ coincides with the electron density, as it is the
case in clean superconductors\cite{tinkham_book}. Indeed, Eq.~\pref{defk} gives:
\be
\lb{defj}
J=22.1 \, n_s^{2D}[10^{13}\mathrm{cm}^{-2}]  \, \mathrm{K}=0.62
\frac{d[A]}{\l^2[(\m{\rm m})]} \, {\rm K}
\ee
With $d$ of order of few nanometers and $\l(0)\sim 100$ \AA, as it is the case
for conventional clean superconductors\cite{tinkham_book}, one has $J_0$
of order of $10^3$ K, so that $t_c$ would be of order of $10^{-3}$, and
then the BKT transition would be essentially indistinguishable from the
mean-field $T_c$.  However, in dirty films of superconductors $J_0$ can be
substantially reduced with respect to the clean case
\cite{fiory_prb83,kapitulnik_prb92,goldman_prb83}, so that $t_c$ assumes
usually values around $0.05 \lesssim t_c \lesssim 0.5$, and $T_{BKT}$ is
sufficiently far from $T_c$ to be observable.

From the point of view of the temperature dependence of the correlation
length one sees that as one moves away from $T_{BKT}$ toward $T_c$ the
decrease of $\mu(T)$ near $T_c$ makes the increase of the vortex fugacity
under RG flow very fast, so that the RG superfluid density vanishes for
$a(\bar \ell)\simeq \xi_0$ and $\xi$ scales as the unrenormalized vortex
fugacity\cite{benfatto_kt_rhos}, i.e.
\be
\lb{xifar}
\xi = \xi_0 e^{-\beta \mu(T)/2}, \quad t\ra t_c.
\ee
Since $\mu(T)\ra 0$ as $T\ra T_c$, it follows that $\xi=\xi_0$ at
$T=T_c$. Notice that strictly speaking one could assume that
also the vortex-lattice spacing $\xi_0$ increases as $T\ra T_c$
as the mean-field correlation length. However, this just
signals that as $T_c$ is approached the phase-modulus
separation that justifies the BKT approach to the fluctuations
fails, because the two degrees of freedom cannot be no
separated anymore in a controlled way. One expects that
$\xi(T)$ interpolates in a continuous way between the BKT
regime \pref{xiapp} and the GL regime, as we shall see below.
Anyway, what is crucial to realize is that the approximate form
\pref{xiapp} is limited to a regime $t\ll t_c$ by the existence
itself of a fluctuating GL regime at higher temperatures.

As it is well know\cite{kosterlitz_renormalization_xy}, in an infinite
system the BKT correlation length remains infinite anywhere below
$T_{BKT}$. This signals the fact that in 2D no symmetry breaking can occur,
so that the SC correlations decay to zero at large distance with a
non-universal power-law decay, instead of the exponential decay to a finite
value that one has in higher dimensions. From the point of view of the
definition of $\xi$ used above in terms of the scale $\ell_s$ where
$J_s(\ell)$ vanishes, since $J_s$ is always finite below $T_{BKT}$ then
$\xi=\infty$ in the SC phase. However, in any real system the RG scaling
must be stopped at a certain scale $\ell_{max}$
\be
\lb{deflmax}
\ell_{max}\equiv \log \frac{L}{\xi_0},
\ee
where $L$ is the maximum physical scale accessible in the system (like the
system size or the size of the homogeneous domains, see Sec.\
\ref{sec:inhom}). As a consequence, the divergence itself of the
correlation length is cut off below the temperature where
$\ell_s=\ell_{max}$, so that the correlation length starts deviating from
Eq.~\pref{xiapp}, and finite-size effects dominate. Sufficiently below
$T_{BKT}$ $K(\ell)$ is substantially unrenormalized with respect to the
initial value $K(0)$,\cite{giamarchi_book_1d} so that we can estimate
finite-size effects by integrating the RG equation for $g$ using
$K(\ell)\approx K(0)$. Thus, $g_u(\ell)=g_u(0)e^{(2-K)\ell}$ from
Eq.~\pref{eqg} and the correlation length behaves as:
\bea
\lb{xiint}
\frac{\xi}{\xi_0}=e^{\beta\mu(T)/2}\left(\frac{L}{\xi_0}\right)^{K(T)/2}, \quad
T \lesssim T_{BKT}. \eea
We notice that, following Eq.~\pref{japp}, near $T_c$, $K(T)$
can be approximated as $K(T)=2(1-t/t_c)$. As a consequence,
from Eq.~\pref{xiint} we see that the size-limited $\xi$ grows
below $T_{BKT}$ as:
\be
\frac{\xi}{\xi_0}\sim \left(\frac{L}{\xi_0}\right)^{\frac{|t|}{t_c}}.
\ee
Thus, if one considers two cases with the same $L$ and $T_{BKT}$ but
different $t_c$ the finite-size effects are less pronounced in the case
with smaller $t_c$, because $\xi$ is still very large below $T_{BKT}$.

\begin{figure}[htb]
\includegraphics[scale=0.3,angle=-90]{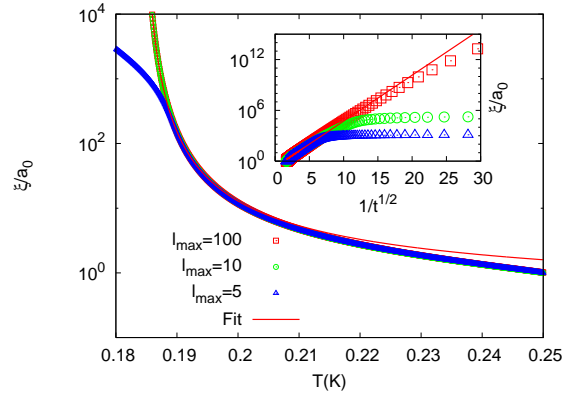}
\caption{(Color online) Temperature dependence of the correlation length
  within KT theory, obtained by numerical integration of the RG equations
  \pref{eqk}-\pref{eqg} with the definition Eq.\ \pref{defnv}, where $\bar
  \ell$ is determined either by the vanishing of the superfluid stiffness
  or by the finite system size $\ell_{max}$, see Eq.\ \pref{deflmax}. Here 
  $T_c=J_0=0.25$ K, and $\tilde \mu=\tilde \mu_{XY}$. The solid line
  represents the fit done using Eq.~\pref{xiapp} with A=5 and
  $b=1.25$. Since $T_{BKT}=0.18$ the $b$ value obtained by the fit is
  consistent with Eq.\ \pref{best}. Inset: same data plotted as a function
  of $1/\sqrt{t}$, where $t$ is the reduced temperature defined in
  Eq.~\pref{red}. Notice that the curves for lower $\ell_{max}$ deviate
  from the infinite-length scale limit \pref{xiapp} at higher
  temperatures.}
\label{fig_xi}
\end{figure}

As an example of the correlation-length temperature dependence we show in
Fig.\ \ref{fig_xi} the numerically-integrated $\xi$ at various values of
the scale $L$, along with the fit of Eq.~\pref{xiapp}. Here we consider the
case $J_0=T_c=0.25$ K, so that a relatively large interval $t_c\simeq 0.36$ is
obtained, to better shown the BKT to GL crossover.  From the inset of Fig.\
\ref{fig_xi} we observe that to correctly capture the exponential scaling
one must use extremely large sizes, far beyond the experimentally
accessible regime. Indeed, since for conventional superconductors usually
$\xi_0\simeq 100$ \AA \ and $L\simeq 1$ cm, then $L=10^6 \xi_0$ and
$\ell_{max}=12$, while the exponential fit in Fig.~\ref{fig_xi} is better
defined with $\ell_{max}=100$.

\begin{figure}[htb]
\includegraphics[scale=0.3,angle=-90]{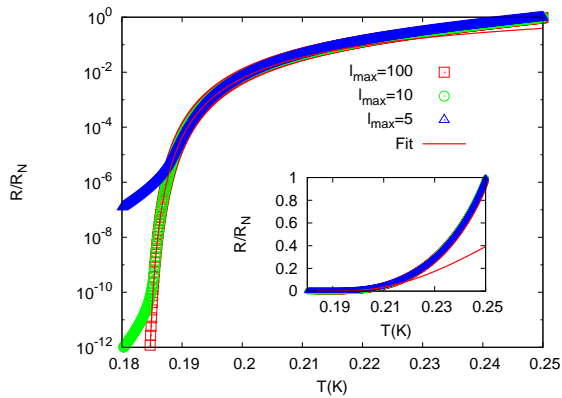}
\caption{(Color online) Fluctuation resistivity in KT theory as a function
  of temperature for several values of the system size. The same set of
  parameters of Fig.\ \ref{fig_xi} has been used. Inset: same curves but on
  a linear scale, where no appreciable difference can be observed for the
  different system sizes.}
\label{fig_res}
\end{figure}

As one can see in the main panel of Fig.~\ref{fig_xi}, the
exponential fit \pref{xiapp} deviates slightly from the RG
correlation length as one moves to temperatures higher than
$T_{BKT}$. This qualitative difference is also reflected by the
temperature dependence of the fluctuations resistivity, that
can be calculated according to Eq.~\pref{rnf}, by using either
the RG or the approximate form \pref{xiapp} of the correlation
length. The result for the same set of parameters used in
Fig.~\ref{fig_xi} is shown in Fig.~\ref{fig_res}: as shown in
the inset, on a linear scale $R$ has always an upward
curvature, with no appreciable differences for the different
$L$ values. Such a difference becomes instead evident on a
logarithmic scale: as $T\ra T_{BKT}$, $R/~R_N$ deviates from the
infinite-size limit, exactly in the same fashion observed in
experiments in thin films (see for example Fig.~9 of
Ref.~\onlinecite{kapitulnik_prb92}). This scaling can be probed
by small magnetic fields: indeed, a small field has the effect
to cut-off the RG at a scale $L_B\propto 1/B$. The curves
shown in Fig.~11 of Ref.~\onlinecite{kapitulnik_prb92} for the
fluctuation resistivity at several (small) values of the
magnetic field reproduce exactly the behavior observed in
Fig.~\ref{fig_res}.

\begin{figure}[htb]
\includegraphics[scale=0.3,angle=-90]{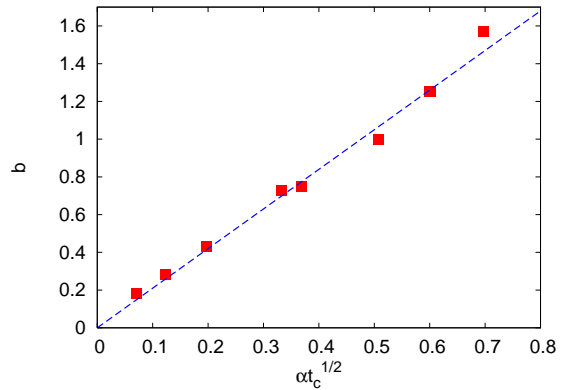}
\caption{(Color online) Estimate of $b$ extracted by fitting the RG
  $\xi(T)$ near $T_{BKT}$ with Eq.~\pref{xiapp} for several values of
  $t_c,\alpha$. The straight line is $y=2.1 x$.}
\label{fig_b}
\end{figure}

Due to the need of exceedingly large system sizes for an
accurate determination of the parameters of the fit
\pref{xiapp}, an a-priori estimate of them as a function of
physical parameters would be particularly useful. To address
this issue we computed $\xi$ for several possible values of
$t_c$ and $\tilde \mu$, to analyze the variations of $b$. While
different $t_c$ values naturally occur in different systems,
the choice of $\tilde \mu$ is somehow still an open problem. As
we mentioned above, within the XY model (that is one of the
possible models for describing phase fluctuations in a
superconductor) $\tilde \mu_{XY}=\pi^2/2$: this value naturally
arises from the mapping of the discrete XY model on the
continuum Coulomb-gas model\cite{nagaosa_book}, and it takes
into account the fluctuations at a scale of the order of the
lattice spacing. In a BCS superconductor one could instead fix
the value of the vortex-core energy by computing exactly the
energy per unit-length of a vortex line\cite{tinkham_book}:
$$
I=\left(\frac{\Phi_0}{4\pi \l}\right)^2 \left [ \log \frac{\l}{\xi_0}+
  \epsilon\right]\equiv \pi J
\left [ \log \frac{\l}{\xi_0}+  \epsilon\right]
$$
so that according to our definition $\tilde \mu=\pi \epsilon$.
A precise estimate of $\epsilon\simeq 0.497$ for the vortex
core in three-dimensional geometry is given in
Ref.~\onlinecite{refmu_hu,refmu_alama}, so that within BCS one
could eventually expect smaller values of $\mu$,
\be
\lb{mu}
\tilde \mu_{BCS}\simeq \frac{\pi}{2}\simeq \frac{\tilde \mu_{XY}}{\pi}.
\ee
Finally, we notice that recent analysis in the context of cuprate
superconductor\cite{benfatto_kt_rhos,benfatto_kt_bilayer,altman_prl07,huse_prb08}
has explored instead the possibility that $\tilde \mu$ is larger than
$\tilde \mu_{XY}$. On the light of the previous observation, we considered
the behavior of the correlation length for a range of values $0.5\lesssim
\tilde \mu\lesssim 1.2$, and we extracted the corresponding $b$ parameter
by fitting data near $T_{BKT}$ with Eq.~\pref{xiapp}.  The results are shown
in Fig.~\ref{fig_b}. As one can see, within a certain degree of
uncertainty, $b$ is found to scale approximately as:
\be
\lb{best}
b\simeq 2\alpha \sqrt{t_c},
\ee
where $\alpha$ is the scale of the vortex-core energy defined in Eq.\
\pref{def_mu_scale}.
Eq.~\pref{best} is the first important result of this paper.
Indeed, it establishes a precise relation between the parameter
$b$ that appears in the typical exponential expression for the
BKT correlation length and the distance $t_c$ between the GL
and BKT temperature. Taking into account
Eq.~\pref{xiapp} and \pref{xifar}, we also see that the formula
\pref{xiapp} can only be used when $b/\sqrt{t}\gg 1$, which
means $t\ll 4 \a^2 t_c$. Indeed, out of this regime $\xi$
decreases according to Eq.~\pref{xifar}, and afterward
($T>T_c$) one enters the GL fluctuations regime. We notice also
that this result agrees with the HN result ~\pref{shn},
once one uses $b_{hn}=2\alpha$.

The analysis we have made by coupling the standard BKT formula \pref{xiapp}
with the RG analysis of the BKT transition thus allows
us to get strong constraints on what the ``fit parameters''
that are to be used in the BKT formula can be. This will help
in the following to take into account additional effects such
as the ones of inhomogeneities, but specially to know if the
fit to the BKT functional form corresponds to a physical fit,
with reasonable parameters, or if one just ``forces the fit'' (see
discussion in Sec.\ \ref{sec:discuss}).

\section{Role of inhomogeneities} \label{sec:inhom}

\begin{figure}[htb]
\includegraphics[scale=0.3,angle=-90]{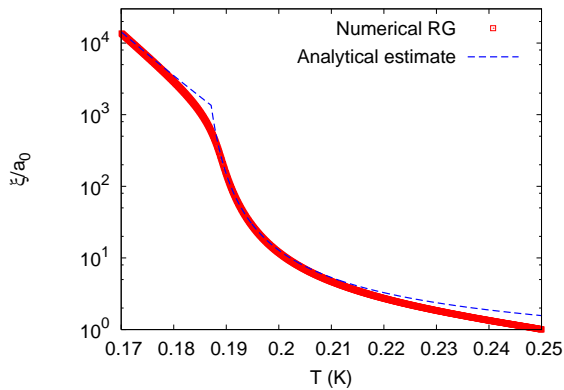}
\caption{(Color online) Comparison between the numerical RG correlation
  length defined by Eq.\ \pref{defnv} and the approximate formula
  \pref{xil}-\pref{xhn}. Here we used the same $T_{BKT},T_c,\mu$ values of
  Fig.\ \ref{fig_xi} in the case $\ell_{max}=5$, so that $b=1.25$ and
  $A=5$.}
\label{fig_study}
\end{figure}

Once we established a clear framework for taking into account
the exponential behavior of the correlation length and the
finite-size effects we can improve the HN original
interpolation formula for paraconductivity, in order to obtain
a self-consistent treatment of SC fluctuations all the way
below and above $T_{BKT}$. We propose the following
interpolating formula:
\be
\lb{int}
\frac{R}{R_N}=\frac{1}{1+({\D \s}/{\s_n})}\equiv \frac{1}{1+(\xi/\xi_0)^2}
\ee
where $\xi$ is given by:
\bea
\lb{xil}
\frac{\xi}{\xi_0}
&=&e^{\beta\mu(T)/2}\left(\frac{L}{\xi_0}\right)^{\pi J(T)/2T}, \quad T
\lesssim T_{BKT} \\
\lb{xhn}
\frac{\xi}{\xi_0}
&=&\frac{2}{A}\sinh\frac{b}{\sqrt{t}}, \quad T \gtrsim T_{BKT}
\eea
with the parameter $A,b$ obtained by the calculated behavior of the numerical
RG correlation length near the transition, so that $b$ is given by Eq.\
\pref{best}, and $A$ is a number of order 1. As it is shown in
Fig.\ \ref{fig_study} this is indeed a very good approximation for the
numerical RG solution near and below $T_{BKT}$. Moreover, even
though it does not capture the regime \pref{xifar}, where the
estimate \pref{xhn} is larger than the RG correlation length,
it correctly reproduces the GL fluctuation regime at $T\gg T_c$,
where $\D \sigma \sim T_c/(T-T_c)$. We stress once more that within this
approach the temperature dependence of the correlation length in the whole
fluctuating regime is uniquely determined by the critical temperatures
$T_{BKT},T_c$ and the value of the vortex-core energy, that can be fixed
by the comparison with the experimental resistivity data. 

\begin{figure}[htb]
\includegraphics[scale=0.3,angle=-90]{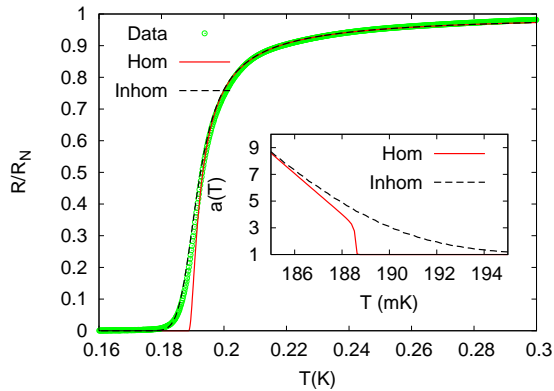}
\caption{(Color online) Comparison between the resistivity of the
  heterostructure measured in Ref.~[\onlinecite{triscone_science07}] and the
  resistivity obtained with the interpolating formula \pref{int}. The curve
  labeled 'Hom' refers to the case of a single local $J$ value, while the
  curve 'Inhom' refers to the resistivity obtained by sample average over
  the local distribution \pref{jdistr} of superfluid density. Inset:
  coefficient $a(T)=1+\pi J_s(T)/T$ of the $I-V$ characteristic in the two
  cases.}
\label{fig_science}
\end{figure}

The main physical message of the set of Eqs.~\pref{best},
\pref{xil}-\pref{xhn} is that from the temperature dependence
of the fluctuation resistivity between $T_{BKT}$ and $T\gg T_c$
one can deduce interesting informations also on the microscopic
parameters of the superconducting system, that determine the
distance between $T_c$ and $T_{BKT}$. To illustrate the
application of this approach we consider the analysis of the
resistivity data in superconducting heterostructures reported
in Ref.~\onlinecite{triscone_science07}. Here the resistive
transition occurs around 0.19 K, but with a relatively large
tail with respect for example to what observed in 2D films of
ordinary
superconductors\cite{fiory_prb83,kapitulnik_prb92,goldman_prb83}.
If we neglect the tail, by means of Eqs.~\pref{xil}-\pref{xhn}
we can obtain the curve labeled as 'Hom' in
Fig.~\ref{fig_science}. As one can see, we reproduce the
overall shape of the resistivity, except from the tail. Since
the fit gives $b=0.19$, according to Eq.~\pref{best} we deduce
that $T_c$ and $T_{BKT}$ almost coincide, with $T_c=0.19$ K and
$T_{BKT}=0.188$ K.

Let us discuss now the origin of the remaining tail of the
resistivity near the transition. First of all, we notice that
it cannot be due to finite-size effects, that can be treated
exactly within our approach. Indeed, even using a relatively
small $L=2$ $\m$m (as suggested by the critical current, see
below), with $\xi_0=70$ \AA\cite{triscone_science07}, one gets
$\ell_{max}=5.6$, but due to the small $t_c$ value the $\xi$
from Eq.~\pref{xil} is still very large below $T_{BKT}$, so the
finite-size effects by themselves are \emph{not responsible}
for the observed tail. On the other hand, from
Fig.~\ref{fig_res} we notice that a tail with upward curvature
is typical of the fluctuation resistivity near the $T_{BKT}$
transition. However since $t_c$ in this case is extremely
small, the tail cannot be even resolved in the scale of
Fig.~\ref{fig_science}. On the contrary, if the transition
\emph{itself} is broadened one could expect to enhance the BKT
tail and to reproduce the experimental data.

On the light of this observation, we suggest that the resistive
tail can be attributed to an inhomogeneous spatial distribution
of the local superfluid, to be ascribed to an intrinsic
inhomogeneous density distribution in these systems. In analogy
with the the analysis of inhomogeneity on the superfluid
density performed in Ref.~\onlinecite{benfatto_kt_bilayer}, we
shall assume for simplicity a gaussian profile:
\be
\lb{jdistr}
P(J)=\frac{1}{\sqrt{2\pi}\d}\exp\left[\frac{-(J-\bar J)^2}{2\d^2}\right]
\ee
where $\bar J$ is the $J_0$ value determined by the above fit
for the homogeneous case. To each $J$ corresponds a mean-field
temperature $T_{mf}=T_c(J/\bar J)$, and accordingly a given
$T_{BKT}$ value. While far from $T_{BKT}$ such an inhomogeneity is
harmless, near $T_{BKT}$ it will give important effects, once we
average over different patches having different transition
temperatures. To compute the effect of the sample inhomogeneity on the
resistivity we go back to Eq.~\pref{rnf}, that
defines the resistivity due to vortices. We notice that the
main quantity that enters the theory is the vortex density:
indeed, the correlation length defined by Eq.~\pref{rnf} is
just a different way to express the vortex density. Thus, in
the presence of the inhomogeneity \pref{jdistr}, we must average
the vortex-density values $n_F(J)$ obtained in each patch with a given
local $J$ value: in this way, even below $T_{BKT}$ there will
be patches of the system where $n_F$ is finite, because $J<\bar
J$ and consequently the local $T_{BKT}$ is smaller than the
average one. As a consequence, the BKT tail gets enhanced, in
excellent agreement with the experiments (with $\delta=0.02
\bar J$), see Fig.~\ref{fig_science}.

A second effect of the inhomogeneity that emerges in the
experiments of Ref.~\onlinecite{triscone_science07} is the lack
of the universal jump of the superfluid density near $J_s$. In
Ref.~\onlinecite{triscone_science07} $J_s$ has been indirectly
measured via the $I-V$ characteristics in the non-homic regime (see
discussion above Eq.\ \pref{icr}), given within the KT
theory by\cite{halperin_ktfilms,review_minnaghen,weber_mc_prb96}:
\be
\lb{eqiv}
V=I^{1+\pi J_s(T)/T}=I^{a(T)}.
\ee
Thanks to the universal relation \pref{jump}, right below the transition
$\pi J_s(T_{BKT})/T_{BKT}=2$, so that the coefficient $a=3$. On the other
hand, in the inifite-size and homogeneous case $J_s(T)$ jumps
discontinuously to zero right above $T_{BKT}$, so that the coefficient
$a(T)$ is expected to jump discontinuously to $1$. Even though
finite-size effects remove the jump and lead to a rapid downturn (see inset
of Fig.~\ref{fig_science}), the measured temperature dependence of $a(T)$
is still much smoother.  This effect can be explained by the presence of
inhomogeneity, that leads to a smooth downturn of $J_s$ near
$T_{BKT}$\cite{benfatto_kt_bilayer}.

As a further check of the correctness of our analysis of the data, we
discuss the implication of the interpolated GL+BKT fit on the estimated
value of the superfluid-density value at $T=0$. From the value of
$t_c\approx 8T_c/(\pi^3 J_0)=0.0106$ obtained by the fit, it follows (see
Eq.~\pref{tcest}) that $J_0=7$ K.  According to Eq.~\pref{defk} above,
$J_0$ is in turn controlled by the zero-temperature value of the density of
superfluid electrons. While in a clean superconductor $n_s^{2D}=n^{2D}$ at
$T=0$, where $n^{2D}$ is the sheet carrier density of the sample, in the
dirty case only a fraction of the order of $\ell_d/\xi_0$ of the electron
density condenses into the superfluid fraction at $T=0$, where $\ell_d=v_F
\tau$ is the mean-free path, with $v_F$ Fermi velocity and $\tau$
scattering time.  Since the BCS correlation length at $T=0$ is $\xi_0=\hbar
v_F/\pi \D(0)$, with $\Delta$ superconducting gap, and $\tau$ can be
determined by the sheet resistance $R_\Box=n^{2D}e^2 \tau/m^*$ (here $\Box$
denotes the sheet area), we obtain an estimate of the disorder-reduced
superfluid density as:
\bea
k_B J_0&=&\frac{\hbar^2 n_s^{2D}}{4 m^*}\approx  \frac{\hbar^2 n^{2D}}{4
  m^*}\frac{\pi \D(0) \ell}{\hbar v_F}=\nn\\
\lb{jest}
&=&\frac{\hbar}{e^2}\frac{ n^{2D}e^2 \tau}{ m^*}\frac{\pi \D(0)}{4}=
\frac{R_c}{R_\Box}\frac{\pi \D(0)}{4}.
\eea
Using the universal value $R_c=\hbar/e^2=4114 \O/\Box$, the measured
normal-state resistance $R\simeq 390 \O/\Box$, and the BCS estimate
$\D(0)=1.76 T_c$ with $T_c=0.19$ K, we obtain $J_0\simeq $ 3 K, which is in
excellent agreement with the value $J_0=7$ K obtained by the fit, taking
into account the approximate estimate of the condensate fraction used in
Eq.\ \pref{jest} above.

\begin{figure}[htb]
\includegraphics[scale=0.3,angle=-90]{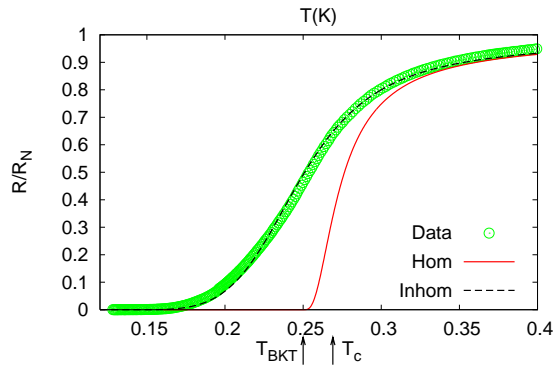}
\caption{(Color online)
Comparison between the experimental data in the unbiased sample (V=0) of
 Ref.~[\onlinecite{triscone_nature08}] and the resistivity obtained with the
  interpolating formula \pref{int}, with ('Inhom') and without ('Hom')
  inhomogeneity. The parameter values are given in the text. The two
  arrows mark $T_{BKT}$ and $T_c$ obtained with the fit in the homogeneous
  case.}
\label{fig_nature}
\end{figure}

The crucial role of inhomogeneities in controlling the tails of
the resistive transition is made even more evident by extending
the above analysis to a second example of superconducting
heterostructures measured in
Ref.~\onlinecite{triscone_nature08}, where the normal-state
resistance is higher, so that inhomogeneous effects could be
expected to be stronger. We shall consider the case of zero
gate voltage in Ref.~\onlinecite{triscone_nature08}, i.e. the
as grown sample without the field-induced doping. The data are
shown in Fig.~\ref{fig_nature} along with our proposed fit
($T_c=0.269$ K, $T_{BKT}=0.25$ K, $b=0.63$ and $J_0=1.05$ K).
Notice that a larger $b$ is consistent with a larger $t_c$,
that in turn can be attributed to a larger disorder in this
second case. Indeed, in this sample $R_\Box=2400$ $\O/\Box$,
and since $n^{2D}=4.5\times 10^{13}$ cm$^{-2}$ using
Eq.~\pref{jest} we obtain a consistently smaller estimate of
$J_0\approx 0.6$ K, which is exactly the result of the fit (a
factor 10 smaller than the previous case). Moreover, the
inhomogeneity is also larger, with $\delta=0.14 \bar J$, and
this explains the very large tail of the transition.

\begin{figure}[htb]
\includegraphics[scale=0.3,angle=0]{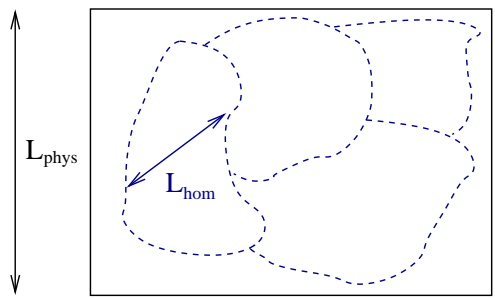}
\caption{(Color online) Schematic sketch of the mesoscopic structure of a
  2D film that can account for the inhomogeneity effects discussed in this
  work. Even though the system does not have a true granular structure the
  homogeneous regions will have a typical size $L_{\rm hom}$ smaller than the
  physical size $L_{\rm phys}$ of the sample. This explains why $L\equiv
  L_{\rm hom}<L_{\rm phys}$ in Eq.\ \pref{icr_est}, giving rise to a critical
  current for linear-to-non-linear characteristic larger than expected for
  the homogeneous case.}
\label{fig_domains}
\end{figure}

As we emphasized above, finite-size effects alone cannot account for the
broadening of the transition. However, one could convert in a hand-waving
way the inhomogeneities effects that we discuss into some form of typical
size for ``homogeneous'' grains in the system. Let us emphasize that this
is just a way to visualize the role of the inhomogeneities and that we do
not assume any sharp granular structure here, see Fig.\ \ref{fig_domains}.
The inhomogeneity distribution (\ref{jdistr}), and in particular the
broadening of $J$ values, $\delta$, could be converted into a
characteristic size $L_{\rm hom}$ much smaller than the true system size.
An indication in this sense is provided by the analysis of the $I-V$
characteristics reported in Ref.~\cite{triscone_science07}. Indeed, within
BKT physics the $I-V$ characteristic is ohmic at low current and non-homic
when the external current is sufficiently high to dissociate
vortex-antivortex pairs. Below $T_{BKT}$, the critical current $I^*$ above
which current-induced free vortices lead to the anomalous power-law
dependence described by Eq.\ \pref{eqiv} above, is in turn related to a
characteristic length scale $L$ where finite-size effects become
relevant\cite{halperin_ktfilms,review_minnaghen,weber_mc_prb96}, so that:
\be
\lb{icr}
I^*=2\pi J_s\frac{c}{\Phi_0}\frac{L_{\rm phys}}{L}
\ee
where $L_{\rm phys}$ is the physical dimension of the sample. 
By using the approximate relation between $J_s$ and
$T_{BKT}$, one can estimate for the sample of
Ref.~\cite{triscone_science07} that
\be
\lb{icr_est}
I^*\simeq
4 K_B T_{BKT}\frac{c}{\Phi_0}\frac{L_{\rm phys}}{L}=\frac{L_{\rm phys}}{L}
0.5 \times 10^{-8}
A
\ee
Since experimentally the critical current is of order of $0.5 \times
10^{-6}$ A and $L_{\rm phys}\simeq 0.2$ mm, we deduce that $L\sim
2\m$m. This can be in turn identified with the size $L_{\rm hom}$ of the
homogeneous domains, so that $\ell_{max}=5.6$, as we used in our
calculations. Notice that Eq.~\pref{icr} accounts also for the temperature
dependence of $I^*(T)$ observed in the experiments: indeed, as we have seen
$J_s(T)$ increases rapidly below $T_{BKT}$ (see the behavior of the $a(T)$
exponent in Fig.\ 5), leading to a very fast increase of the critical
current below the $T_{BKT}$ transition. 
At temperatures well below
$T_{BKT}$ one should further account in the 
analysis of the $I-V$ characteristic for the nucleation of virtual
vortex-antivortex pairs, formed by a single vortex and its virtual image
due to the edge currents in a finite-size system, as pointed out recently
in Ref.~[\onlinecite{gurevich_nucleation_XY}].

\section{Discussion on the analysis of the experimental data}
\label{sec:discuss}

In the previous Section we performed a systematic analysis of
the resistivity data from
Refs.~\onlinecite{triscone_science07,triscone_nature08}. Our
analysis is based on the regular BKT transition and its connection
with Gaussian fluctuations leads to the conclusion that the BKT
transition occurs \emph{very near} (within few percent) of the
mean-field GL transition temperature $T_c$. In this case the
BKT fluctuation regime is restricted to a small range $t_c$ of
reduced temperatures near $T_{BKT}$, where the correlation
length \pref{xhn} has the exponential behavior \pref{xiapp},
with an exponential divergence near $T_{BKT}$ that is
controlled by the same parameter $b$ \pref{best} that measures
the distance from the GL transition $T_c$. At temperatures near
and below $T_{BKT}$ inhomogeneities are responsible for the
tail in the resistive transition.

\begin{figure}[htb]
\includegraphics[scale=0.3,angle=-90]{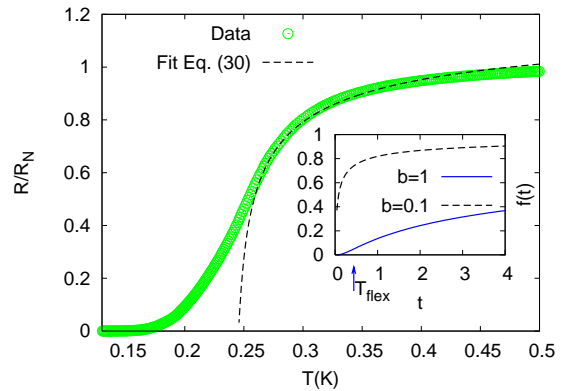}
\caption{(Color online) Example of application of the the BKT fit proposed
  in Ref.\ [\onlinecite{schneider_finitesize09}] for the same data shown in
  Fig.\ \ref{fig_nature}. The theoretical curve is Eq.\ \pref{rapp} with
  $T_{BKT}=0.245$ K and $b=0.106$. Inset: temperature dependence of the
  function $f(t)=\exp(-2b/\sqrt{t})$ for two $b$ values. For $b=1$ one can
  clearly distinguish on this scale a first regime with upward curvature,
  up to the temperature $T_{flex}$ given by Eq.\ \pref{tflex}, followed by
  a regime with downward curvature. For $b=0.1$ only the second regime is
  visible on this scale.}
\label{fig_schneider}
\end{figure}

A quite different way to interpret the data was followed
instead in
Refs.~\cite{triscone_science07,triscone_nature08,schneider_finitesize09},
based essentially on the idea that $T_c$ is far larger than
$T_{BKT}$, so that the \emph{whole} fluctuation regime should
be described via the standard BKT approach, see Fig.\
\ref{fig_scheme}. For the sake of clarity in the following we will refer
to our approach as 'BKT-GL'(Fig.\ \ref{fig_scheme}a), and to the one 
proposed in Refs.\
\cite{triscone_science07,triscone_nature08,schneider_finitesize09} as
'BKT-only' (Fig.\ \ref{fig_scheme}b). In the latter case, the
expression \pref{xiapp} for the correlation length should be valid in
\emph{all} the temperature regime where $R/R_N$ deviates from
$1$, and $b$ is a number of order $1$, to assure that $T_c$ is
enough far from $T_{BKT}$. As a consequence, the \emph{whole}
resistivity \pref{rnf} is given at all temperature above
$T_{BKT}$ by:
\be
\lb{rapp}
\frac{R}{R_N}=A^2e^{-2b\sqrt{T_{BKT}/({T-T_{BKT}})}}=A^2 f(t),
\ee
where $f(t)=\exp(-2b/\sqrt{t})$ and $t$ is the reduced temperature \pref{red}.
An example of application of Eq.\ \pref{rapp} to the same resistivity data
analyzed in Fig.\ \ref{fig_nature} above is presented in Fig.\
\ref{fig_schneider}. Here we used Eq.\ \pref{rapp} with $T_{BKT}=0.245$ K
and considered for the moment $b$ as a completely {\em free} parameter. In
this case, a reasonable fit of the data can be obtained by using
$b=0.106$. Such a small value of $b$ can be understood by looking at the
behavior of the function $f(t)$ shown in the inset of Fig.\
\ref{fig_schneider}. As one can notice, if one extend $f(t)$ at arbitrary
$t$ values above $T_{BKT}$ it actually saturates to a constant
value. The crossover from the low $T$ regime where $f(t)$  displays
an upward curvature to the one where it displays an downward
curvature can be estimated by the flex of this curve, that is
found at:
\be
\lb{tflex}
T_{flex}=T_{BKT}[1+0.44 b^2].
\ee
As a consequence, if $b$ is very small the function $f(t)$ saturates very
rapidly and one can fit the experimental data with the expression
\pref{rapp} by assuming a very small $b$ value. However, this procedure is
totally inconsistent with the fact that within BKT theory $b$ is not a free
parameter, as we discussed in the previous Sections, but it depends,
through Eq.\ \pref{best}, on the distance $t_c$ between the BKT and GL
temperatures. In particular, the fit presented in Fig.\ \ref{fig_schneider}
gives $b=0.1$, that would imply a $t_c=0.0025$, in clear
contradiction with the a-priori assumption that $T_c$ is so much larger
than $T_{BKT}$ than the whole fluctuation regime has BKT character.
Moreover, a smaller $b$ value would also imply that $J_0$ is very large
($J_0\sim 26$ K), that is again inconsistent with the estimate \pref{jest}
based on the normal-state resistivity value (i.e. $J_0\sim 7$ K).  We note
in particular that this procedure leads to a value for the parameter $b$
that is about {\em one order of magnitude smaller} than what we obtained in
our case (see Fig.\ \ref{fig_nature}), where $b=0.6$. Indeed, in the BKT-GL
approach most of the fluctuation regime is accounted by GL fluctuations, so
that Eq.\ \pref{rapp} is actually valid only in the limited range of
temperatures $T\ll T_{flex}$ where $f(t)$ displays an upward curvature.

An incorrect use of the BKT formulas can lead to quite
unphysical values for the $b$ parameter, as it is the case in our opinion
for Ref.~[\onlinecite{schneider_finitesize09}].  
Moreover, 
independently of the question of the distance between $T_{BKT}$ and
$T_{c}$, a second drawback of this BKT-only analysis is that it attributes 
the tail of the transition, that is the real signature of BKT physics, to
unrealistic finite-size effect. Indeed, in
Ref.~[\onlinecite{schneider_finitesize09}] 
it has been proposed that the deviation from the best fit obtained with
Eq.~\pref{rapp} and the experimental data occurs when $\xi$ is cut-off by
finite-size effects. As we have seen above, finite-size effects become
relevant near $T_{BKT}$ when $\xi\sim L$, see Eq.\ \pref{xiint}. Thus,
since $R/R_N= (\xi_0/\xi)^2$ according to Eq.\ \pref{rnf}, in  
Ref.~[\onlinecite{schneider_finitesize09}] it has been argued that the fit
with Eq.\ \pref{rapp} deviates from the experimental data when
$\xi(T)=L$, with $T>T_{BKT}$. By applying this idea, since from Fig.\
\ref{fig_schneider} it follows that the fit deviates from the data already
at $R/R_N\sim 0.5$, one would obtain $L\sim 2 \xi_0\sim 100$ \AA. This is an
unrealistic small number for the size of the homogeneous domains, because
it would not allow at all the formation of coherent vortex structures (and
then the observation of their signatures) on a distance as small as only
twice the lattice spacing for phase fluctuations (i.e. $\xi_0$). In our
approach instead the deviation of the experimental data from the
homogeneous-case fit is due to inhomogeneity, while finite-size effects
alone cannot account for it.

The only way to possibly justify the BKT-only analysis of the data would be
if for some reason the constraints on the $b$ parameter that we establish
in our paper for a conventional BKT transition would not be obeyed, namely
if the transition is of a different nature than the conventional BKT
transition. What could be such an alternative is unclear
from a theoretical point of view. A proposal in this sense has been
formulated in Ref.~[\onlinecite{triscone_science07}], where it has been
argued that in these systems one can estimate the initial value of the
vortex fugacity in the RG equations \pref{eqk}-\pref{eqg} to be quite
large, invalidating the applicability itself of the perturbative
RG\cite{review_minnaghen,gabay_melting}. In this case, following the
analysis of Ref.\ [\onlinecite{gabay_melting}], the BKT transition has a
qualitative different character, since it refers to the melting of the 2D
lattice of vortices formed at high vortex density.  Such a scenario can
provide for example an alternative interpretation to the high values of the
critical current $I^*$ \pref{icr} mentioned above (see Supplementary
Information of Ref.\ [\onlinecite{triscone_science07}]).  This is thus an
interesting proposal, but it leads to a BKT-only analysis of the data that
still suffers from two main problems: a) No theoretical work has been
devoted so far to explore the signatures of such a melting transition, if
it exists, on the SC fluctuations contribution to the resistivity. In
particular it is not at all obvious that such a scenario would lead to a
different $b$ parameter, so that the small $b$ value obtained by the
application of Eq.~\pref{rapp} at all temperatures above $T_{BKT}$ would be
meaningful. b) Recent numerical simulations on the vortex liquid do not
confirm the existence of this crystal
phase\cite{lidmar_bkt_simulation_prb97} at even at high vortex fugacity and
only standard BKT vortex phase seems to be present, so that it is not even 
clear whether such a melting transition would exist.  Thus, more
theoretical and experimental work is certainly needed to test this
possibility and to decide which one of the interpretations, the BKT-GL
approach or the BKT-only based on the idea of vortex lattice melting, is
indeed the correct explanation of the existing data. For example, further
measurements of the size-dependence of the critical current $I^*$ or the
direct measurement (via two-coils mutual inductance experiments) of the
temperature dependence of the superfluid density can provide us with
additional information on the system, to be tested against the different
theoretical proposals.

\section{Conclusions} 
\label{sec:concl}

In the present work we investigated the role of finite-size effects and
inhomogeneity of the BKT transition in quasi-2D SC systems. By analytical
and numerical analysis of the RG equations for the BKT transitions we
determined in a precise way how the properties of the system control the
behavior of the correlation length above and below the transition. By
interpolating the BKT behavior with the GL fluctuations we propose an
unified scheme to determine both the mean-field transition temperature
$T_c$ and the BKT one $T_{BKT}$ from a fit of the resistivity data. In our
approach, those that are usually treated as free parameters in the BKT
fitting formula are implicitly determined by the these temperature scales,
and allows us to identify the effects that must be attributed to the finite
system size and inhomogeneity.  The direct comparison with recent
experiments in SC heterostructures suggests that spatial inhomogeneity,
whose role we discussed already in the context of ultra-thin films of
high-$T_c$ superconductors\cite{benfatto_kt_bilayer}, is a common
ingredient to these systems as well. Our analysis allows us to estimate
also the superfluid-density content of these unconventional SC interfaces,
whose direct experimental determination is not yet available.
Interestingly, in Ref.~[\onlinecite{triscone_nature08}] it has been shown
that by decreasing the carrier density by field-effect one can induce in
such heterostructures the transition from a SC to a metallic and then
insulating state.  An interesting open issue is the role played by
inhomogeneity in such a crossover: for example, one could expect that as
$J_0 (T_c)$ decreases and $\delta$ increases one reaches the percolative
threshold below which the system cannot sustain anymore superfluid
currents, in analogy with the behavior of diluted XY
models\cite{pereira_percolation_prb05}. This issue can have profound
consequences on the nature of the quantum critical point inferred in
Ref.~[\onlinecite{triscone_nature08}], and certainly deserves further
investigation.

\section{Acknowledgments}
We thank J.~Lesueur for useful discussions and suggestions. We 
thank A.~Caviglia, M.~Gabay, S.~Gariglio, N.~Reyren and J.~M.~Triscone for
several stimulating discussions and for providing us with the data of
Refs.~[\onlinecite{triscone_science07,triscone_nature08}], shown in
Figs.~\ref{fig_science},\ref{fig_nature}.  T.~G.~ thanks the KITP, where
part of this paper was completed for hospitality and support under the NSF
grant PHY05-51164.  This work has been supported in part by the Swiss
National Science Foundation under MaNEP and Division II and by the Italian
MIUR under the project PRIN 2007FW3MJX.

\bibliographystyle{apsrev}


\end{document}